\documentclass{PoS}

\usepackage{subfig}

\title{$F_K/F_\pi$ in full QCD}

\ShortTitle{$F_K/F_\pi$ in full QCD}

\author{\speaker{A. Ramos}\\
  Centre de Physique Th\'eorique. CNRS Luminy, Case 907,
  F-13288 Marseille Cedex 9, France.\\ 
  E-mail: \email{Alberto.Ramos@cpt.univ-mrs.fr}} 

\author{For the Budapest-Marseille-Wuppertal Collaboration}

\abstract{We determine the ratio $F_K/F_\pi$ in QCD with $N_f=2+1$ flavors of
sea quarks, based on a series of lattice calculations with three
different couplings, large volumes and a simulated pion mass reaching
down to about 190\,MeV. We obtain 
$F_K/F_\pi = 1.192(7)_{stat}(6)_{sys}$ with all the sources of
systematic uncertainty under control.}

\FullConference{The XXVII International Symposium on Lattice Field Theory - LAT2009\\
		 July 26-31 2009\\
		 Peking University, Beijing, China}

\begin{document}

\newcommand{\bdm}{\begin{displaymath}}
\newcommand{\edm}{\end{displaymath}}
\newcommand{\beq}{\begin{equation}}
\newcommand{\eeq}{\end{equation}}
\newcommand{\bea}{\begin{eqnarray}}
\newcommand{\eea}{\end{eqnarray}}

\newcommand{\Mpi}{M_\pi}
\newcommand{\Fpi}{F_\pi}
\newcommand{\Mka}{M_K}
\newcommand{\Fka}{F_K}
\newcommand{\Met}{M_\eta}
\newcommand{\Fet}{F_\eta}

\newcommand{\mr}{\mathrm}
\newcommand{\ovr}{\over}
\newcommand{\fm}{\,\mr{fm}}

\def\reff#1{\ref{#1}}
\def\labell#1{\label{#1}}
\def\app#1{Appendix~\reff{#1}}
\def\apps#1#2{Appendices~\reff{#1}--\reff{#2}}
\def\eq#1{Eq.~(\reff{#1})}
\def\eqs#1#2{Eqs.~(\reff{#1})--(\reff{#2})}
\def\fig#1{Fig.~\reff{#1}}
\def\figs#1#2{Figs.~\reff{#1}--\reff{#2}}
\def\sec#1{Sec.~\reff{#1}}
\def\tab#1{Table~\reff{#1}}
\def\tabs#1#2{Tabs.~\reff{#1}--\reff{#2}}

\def\lsim{\raise0.3ex\hbox{$<$\kern-0.75em\raise-1.1ex\hbox{$\sim$}}}
\def\gsim{\raise0.3ex\hbox{$>$\kern-0.75em\raise-1.1ex\hbox{$\sim$}}}

\def\mev{\mathrm{MeV}}

\section{Introduction}

We determine $F_K/F_\pi$ in QCD through a series of
dynamical lattice 
calculations such that all the sources of systematic uncertainty are
properly taken into account~\cite{Durr:2010hr}. We use $N_f=2+1$
dynamical quarks, with a 
single quark whose mass is close to the physical strange quark mass
($m_s\!\simeq\!m_s^\mr{phys}$), and two degenerate flavours of light
quarks heavier than in the real world $u$ and $d$ quarks, but with
masses varying trough a range that allows a controlled extrapolation
to the physical point. Concerning finite volume effects, the
spatial size $L$ is large enough so that the values of $F_K/F_\pi$in
our ensembles can be corrected for small finite-volume effects.
We simulate at three different values of $\beta$ to have full control
over the continuum extrapolation.

\section{Simulation and analysis details}

Here we will not give any details about our actions for the gauge and
fermion fields, or about the algorithms used for the simulation.
The interested reader should consult~\cite{Durr:2008rw}.

To set the scale and adjust the quark masses, we use $aM_\pi$, $aM_K$
and either $aM_\Xi$ or $aM_\Omega$ (to estimate the systematics, as we
will see). We
extrapolate for each value of the lattice spacing the values
$aM_\pi,aM_K,aM_\Xi$ to the point where any two of the ratios agree
with the experimental values. We subtract electromagnetic and isosping
breaking effects to the experimental values of the masses:  we use 
$\Mpi^\mr{phys}=135\,\mev$, $\Mka^\mr{phys}=495\,\mev$ and
$M_\Xi^\mr{phys}=1318\,\mev$ with an error of a few
MeV, according to~\cite{Aubin:2004fs}. 

Regarding $m_{ud}$, we have pion masses in the range $190-460\,
\mev$. $F_K/F_\pi$ is measured with the valence quark masses equal to
the sea quark masses (no partial quenching). 

The same dataset was successfully used to
determine the light hadron spectrum~\cite{Durr:2008zz}.

\section{Treatment of the theoretical errors}

\subsection{Extrapolation to the physical mass point}

We simulate with a strange quark mass already close to its physical
value, but this is not the case for the light quarks. Thus our values 
of $F_K/F_\pi$ need an extrapolation to the physical point. There are
three possible guides for this extrapolation: $SU(3)$ chiral
perturbation theory, heavy kaon $SU(2)$ chiral perturbation theory,
and Taylor fits. 

For the case of $SU(3)$ chiral perturbation theory we use the
expression for the ratio at NLO as a function of the measured pion and
kaon masses. It is important to note here that the apparent
convergence of 
Chiral perturbation theory is a statement that depends both on the
observable and the statistical accuracy of the data. With our data, the
ratio $F_K/F_\pi$ is well described by the NLO expression, but this
does not mean that NLO expressions can be used in general to describe
others quantities (see~\cite{Lellouch:2009fg}). The ratio $F_K/F_\pi$
is probably a benevolent quantity as chiral logs in $F_\pi$ and $F_K$
cancel in part.

$SU(2)$ chiral perturbation theory~\cite{Gasser:1983yg} and its heavy
kaon variant~\cite{Allton:2008pn} determines the functional dependence
of the ratio of decay constants on pion mass. We find that the ratio
of NLO expressions from these two references describes well the data
in our ensembles.

Having data in the range $190\,\mev - 460\,\,\mev$, it is natural to 
consider an expansion about a regular point which encompasses 
both the lattice results and the physical
point~\cite{Durr:2008zz,Lellouch:2009fg}. The mass dependence of
$F_K/F_\pi$ in 
our ensemble and at the physical point is well described by a low
order polynomial in $M_\pi^2$ and $M_K^2$.

Since all the three frameworks describe well our data, we will use all
of them in our analysis, and use the difference between them to estimate the
systematic uncertainty.

\subsection{Continuum limit}
\label{sec:contlim}

$F_K/F_\pi$ is an $SU(3)$-flavor breaking ratio, so that cutoff
effects must be proportional to $m_s-m_{ud}$, that guided by 
$SU(3)$ chiral perturbation theory we may substitute with
$M_K^2-M_\pi^2$. 

Cutoff effects
are both theoretically (they partially cancel in the ratio) and in practice
small numerically.  In \cite{Durr:2008rw}
we found that although our action is only formally improved up to
order $a$, these small cutoff effects seem to scale like $a^2$ up to
about 0.16~fm. Even if this is the case we can not exclude the
possibility of cutoff effects proportional to $a$. Thus we have
considered the following three options compatible with our data: no
cutoff effects, and a flavor breaking term proportional to $a$ or $a^2$.

\subsection{Infinite volume limit}

Stable states in a box with periodic boundary conditions have
different masses and decay constants than the corresponding states in
infinite volume. The difference vanishes exponentially fast with the
mass of the lightest state in the box~\cite{Luscher:1985dn}. In our
case, masses and decay constants are corrected with terms proportional
to $\exp(-\Mpi L)$. In our simulations $\Mpi L\!\gsim\!4$ making finite
volume corrections small. Moreover the sign of leading correction in
$F_\pi$ and $F_K$ is the same, so that they partially cancel in the ratio.

Within chiral perturbation theory, the
1-loop~\cite{Gasser:1986vb,Becirevic:2003wk} and
2-loop~\cite{Colangelo:2005gd} corrections for the pion and kaon decay
constants are known, so we have decided to correct the values of our
simulations with the 2-loop expression before fitting the data. The
1-loop expression will be used to estimate the systematic uncertainty
due to finite volume corrections (see below).

Similarly, we also correct the meson masses~\cite{Colangelo:2005gd}. 

\section{Fitting strategy and treatment of theoretical errors}
\label{sec:fitting}

Our goal is to obtain $F_K/F_\pi$ at the physical point, in the
continuum and in infinite volume.  To this end we perform a global fit
which simultaneously extrapolates or interpolates
$\Mpi^2\to\Mpi^2|_\mr{phys}$, $\Mka^2\to\Mka^2|_\mr{phys}$ and
$a\to0$, after the data have been corrected for very small finite
volume effects using the two-loop chiral perturbation theory results
discussed above. To assess the various systematic uncertainties
associated with our analysis, we perform a large number of alternative
fits.

Excited states contribute to the correlators, so to estimate their
effect, we choose 18 different time intervals dominated by the ground
state. The difference between masses and decay constants coming from
different time intervals are used to estimate the uncertainty
associated with excited states.

Scale setting systematic uncertainty is estimated by using 
$M_\pi$, $M_K$ and either $M_\Xi$ or $M_\Omega$.

The chiral extrapolation systematic error is estimated in two
ways. First we consider two different ranges of pion masses for the
fits: 
$190-350\, \mev$ and $190-460\, \mev$. Second, we consider a total of
7 different functional forms to extrapolate to the physical point. 3
of them come from the NLO $SU(3)$ chiral perturbation expression:
the ratio of decay constants (cancelling terms proportional to $L_4$),
a NLO expansion of the ratio and a similar NLO expansion but for the
inverse ratio $F_\pi/F_K$. 2 functional forms come from the $SU(2)$ chiral
perturbation theory expression: the expanded ratio of decay constants,
and a similar expression but for $F_\pi/F_K$. The last 2 forms
correspond to a Taylor fit for $F_K/F_\pi$, and a Taylor fit for
$F_\pi/F_K$. It is important to note that the difference between these
7 functional forms provides an estimate of both higher order contributions
within a concrete framework (e.g NNLO terms in $SU(3)$ chiral
perturbation theory), and systematic bias coming from a particular
framework. 

As discussed above, cutoff effects are parametrised in three different
ways:  we consider fits with and without $O(a^2)$ and $O(a)$
corrections, as described in \sec{sec:contlim}.

\begin{figure}[t,b]
  \centering \subfloat[Chiral extrapolation of the lattice data to the
  physical point for a particular choice of time interval and mass cut
  ($M_\pi<460\,\mev$ in this case). Here we use a Taylor ansatze with
  no cutoff effects. The $m_s$ dependence has been
  subtracted to plot the data as a function only of $M_\pi$.
  ]{\label{fig:fit}\includegraphics[width=7.3cm]{fig/fkovfpi}}\quad
  \subfloat[Distribution of values of $F_K/F_\pi$. The large
  background distribution (yellow) represents the values of
  $F_K/F_\pi$ obtained with different extrapolation formulas, pion
  mass cuts, parameterization of cutoff effects, time intervals and
  different methods to set the scale. Also shown is the final
  error interval (dashed lines) and the final value (black
  solid vertical
  line).]{\label{fig:dist}\includegraphics[width=7.3cm]{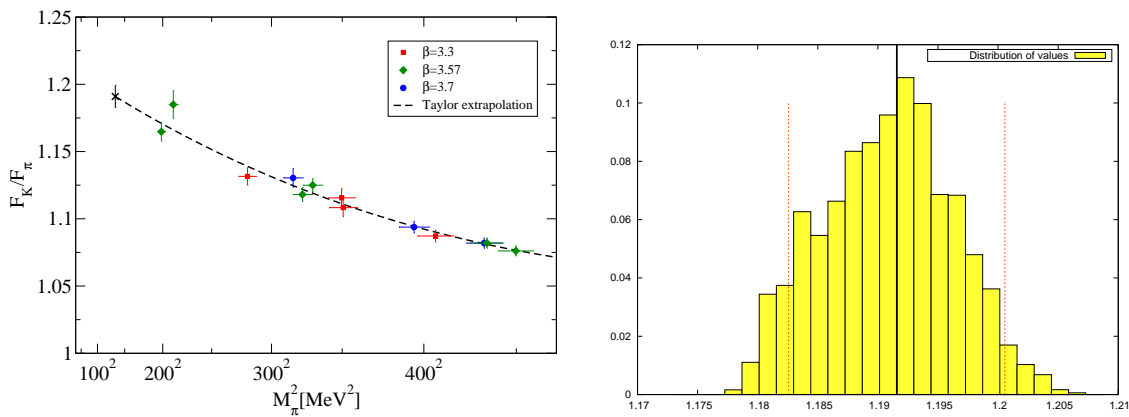}}
  \caption{And example of mass extrapolation and the distribution of
    fits used to obtain the final result and systematic error.}
\end{figure}

Following this procedure we have $18\times 2\times 2\times 7\times 3=1512$
global fits. One of the 1512 fits (corresponding to a specific choice
for the time intervals used in fitting the correlators, scale setting,
pion mass range, \dots) can be seen in~(\fig{fig:fit}). We emphasize that
the $\chi^2$ per d.o.f. for our correlated fits are close to
one. 

Although all these methods seem to describe well our
data, not all of them do it in the same way, so we weight with the fit
quality the central value of each fit. These 1512 weighted values can
be used to construct a distribution, 
whose median is an estimate of the typical result of our analysis,
therefore our desired final result  (see fig.~\ref{fig:dist}). The
width of the distribution is a  
measure of the systematic error of our analysis, thus we take the
16-th/84-th percentiles as an estimate of the systematic error of our
computation.

Finite volume effects are treated separately because we know a priori
that the two-loop expressions of \cite{Colangelo:2005gd} are the most
accurate expressions available and they describe well these effects in
our data. To estimate the error associated with the finite volume
effects, we repeat the full analysis using the 1 loop expression to
correct the ratio $F_K/F_\pi$ and we also repeat the full analysis
correcting only the value of $F_\pi$ (this can
be seen as an upper bound to the real correction in $F_K/F_\pi$). The
weighted (with the quality of the fit) standard deviation of these
three values is used as an estimation of the uncertainty due to finite
volume effect, and added to our systematic error by quadratures to
produce the final systematic error.

To determine the statistical error, the whole procedure is
bootstrapped with 2000 samples, and the standard deviations of the
2000 medians used as our estimate of the statistical error. Our final
error is computed as the sum by quadratures of the total systematic
and statistical errors. 

\section{Results}

Our final result for the ratio of decay constants is
\beq
{F_K\over F_\pi}\bigg|_\mr{phys}=1.192(7)_\mr{stat}(6)_\mr{syst}
\qquad\mbox{or}\qquad
{F_\pi\over F_K}\bigg|_\mr{phys}=0.839(5)_\mr{stat}(4)_\mr{syst}
\label{final_ratio}
\eeq
at the physical point, where all sources of systematic error have
been included. 

Figure~\ref{fig:comp} shows  our final result compared with the
determination of $F_K/F_\pi$ from other dynamical lattice
computations. There are two $N_f=2$ computations by JLQCD 
\cite{Aoki:2002uc} and ETM \cite{Blossier:2009bx}. With $2+1$ fermion
flavours, 
we have a number of results obtained using MILC configurations: MILC
\cite{Aubin:2004fs,Bazavov:2009bb}, NPLQCD \cite{Beane:2006kx}, 
HPQCD/UKQCD \cite{Follana:2007uv} and Aubin et al.\
\cite{Aubin:2008ie}. The results by RBC/UKQCD \cite{Allton:2008pn} and
PACS-CS \cite{Aoki:2008sm} were also obtained with $N_f=2+1$
simulations but with different configurations.  It is worth noting
that these results show a good overall consistency when one excludes
the outlier point of \cite{Aoki:2002uc}. 

\section{Contributions to the systematic error}

Having estimated the total systematic error, it is interesting to
decompose it into its individual contributions. To quantify these
contributions, we construct a distribution for each of the possible
alternative procedures corresponding to the source of theoretical
error under investigation. These distributions are constructed by
varying over all of the other procedures and weighing the results by
the total fit quality. Then, we take the weighted standard deviations
of the medians of these distributions as an estimate of the systematic 
uncertainty associated with the source of error under consideration. 

\begin{figure}[t,b]
  \centering
  \subfloat[Comparison of recent lattice computations of $F_K/F_\pi$.]{\label{fig:comp}\includegraphics[width=7cm]{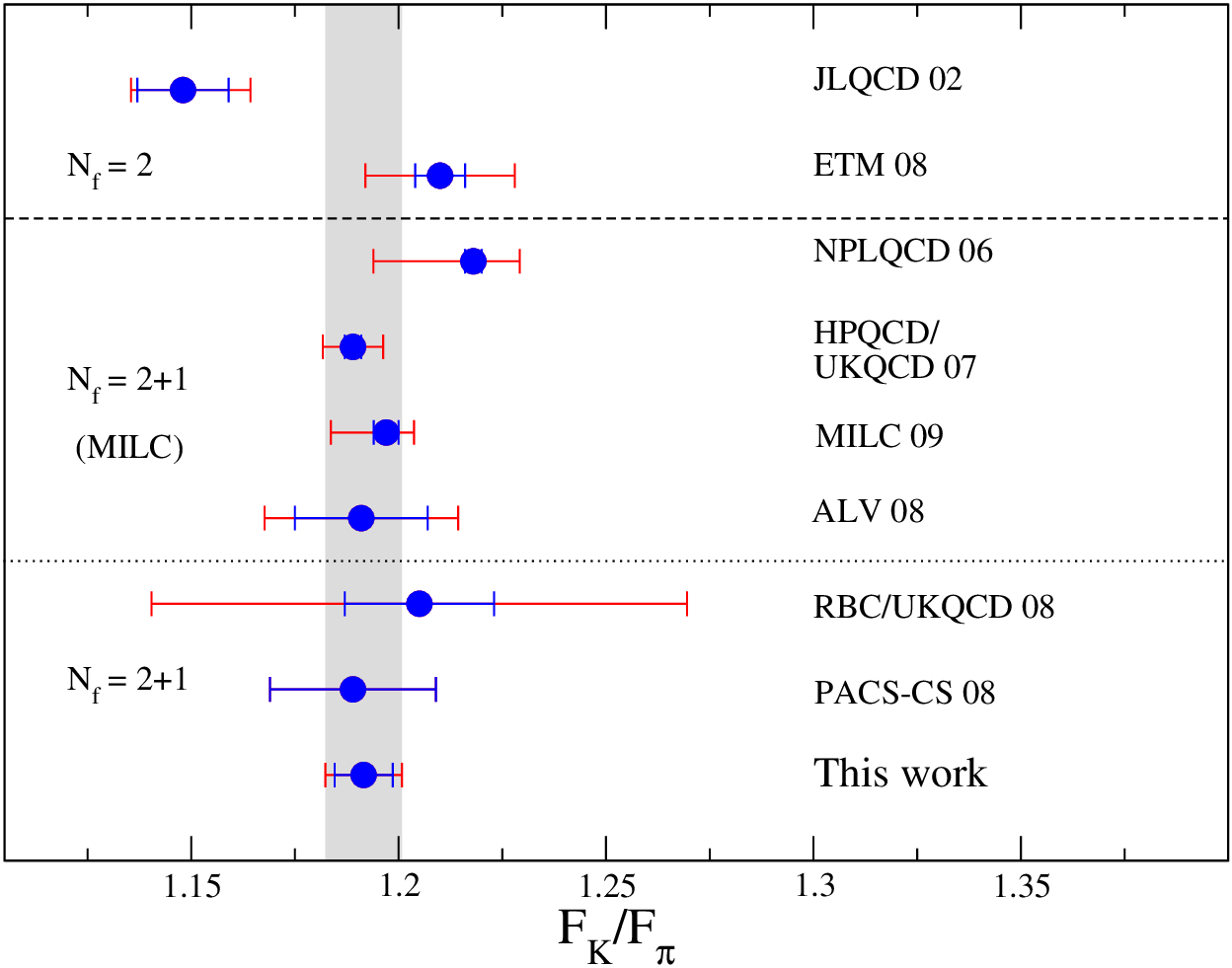}}\quad
  \subfloat[Breakdown of the total systematic error on $F_K/F_\pi$
    into its various components, in order of decreasing
    importance.]{\label{tab:err}\begin{tabular}[b]{lc}
    \hline
    \hline
    Source of systematic error & {error on $F_K/F_\pi$} \\
    \hline
    Chiral Extrapolation: &  \\
     - Functional form & $3.3\times 10^{-3}$ \\
     - Pion mass range & $3.0\times 10^{-3}$ \\
    Continuum extrapolation & $3.3\times 10^{-3}$ \\
    Excited states & $1.9\times 10^{-3}$ \\
    Scale setting & $1.0\times 10^{-3}$\\
    Finite volume & $6.2\times 10^{-4}$\\
    \hline
    \hline
  \end{tabular}} \\ 

  \caption{In the figure, we can see a comparison
    between our result and recent unquenched computations of
    $F_K/F_\pi$. The table shows, in order of
    importance, the different sources of systematic error.}
\end{figure}

\begin{figure}[t,b]
  \centering
  \subfloat[Final distribution of values and the contribution to this
  corresponding to different pion mass cuts.]{\includegraphics[width=7.3cm]{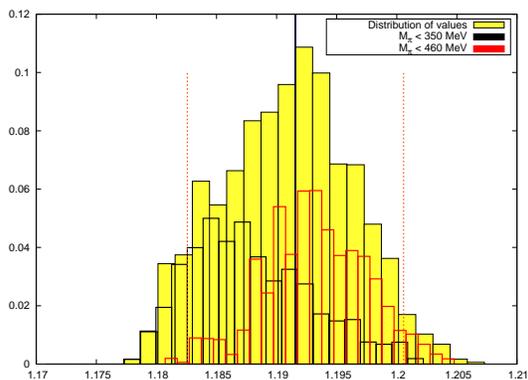}}\quad 
  \subfloat[Final distribution of values and the contribution to this
  corresponding to different functional forms.]{\label{fig:mth}\includegraphics[width=7.3cm]{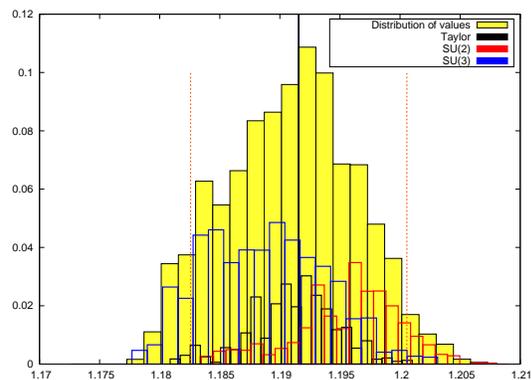}}\\
  \caption{Analysis of chiral extrapolation error.}
\label{fig:ext}
\end{figure}

Table~(\ref{tab:err}) shows the estimation of the different sources of
systematic error in our computation. Even having pion masses down to
$190$~MeV, the chiral extrapolation remains the main source of
systematic error. This error is broken in two parts by changing the
fit range, and using different expressions to extrapolate the data. In
the \fig{fig:ext} we can see how fits corresponding to
different pion mass cuts, and different functional forms contribute to
the final distribution of values. The medians of the small constituent
distributions are used to compute the error associated with each
source of error, as was mentioned before.

\begin{figure}[t,b]
  \centering
  \subfloat[Final distribution of values and the contribution to this
  corresponding to different cutoff effects.]{\label{fig:cont}\includegraphics[width=7.3cm]{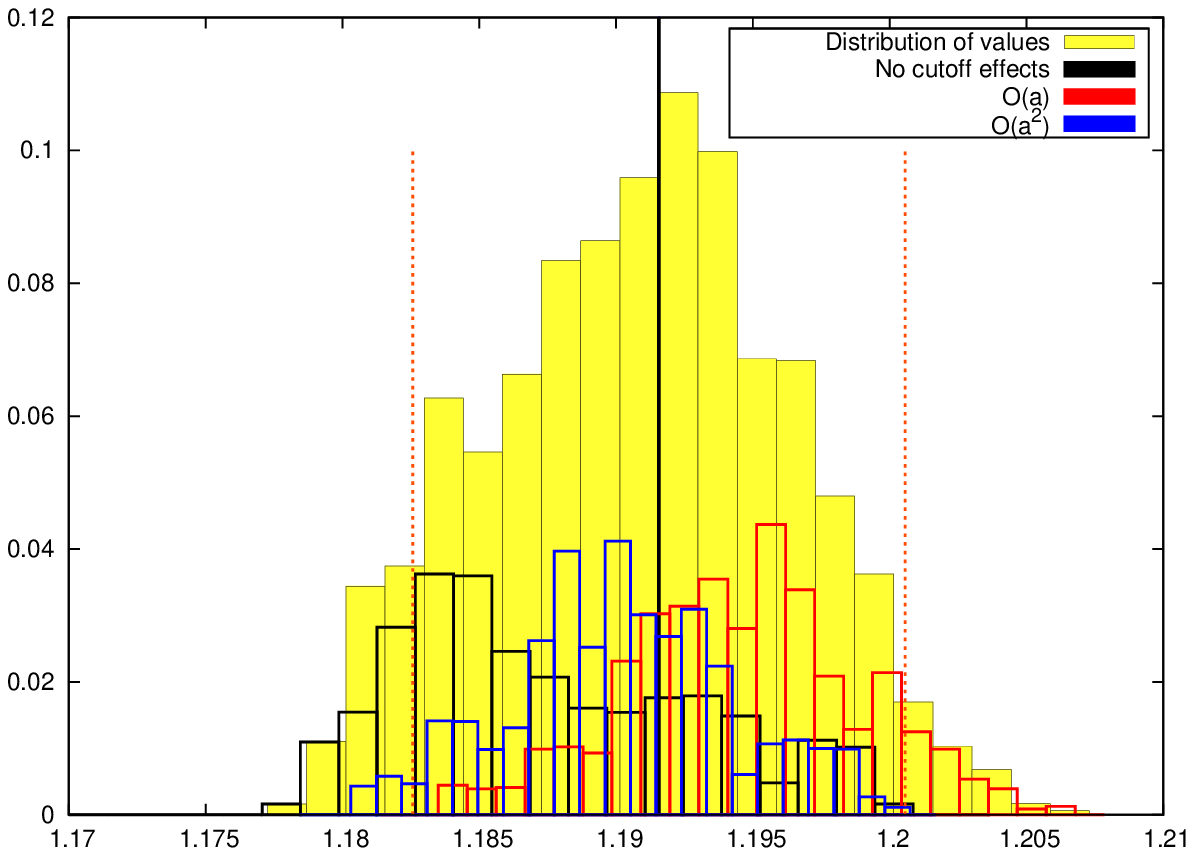}}\quad 
  \subfloat[Final distribution of values and the contribution to this
  corresponding to two different fitting ranges (of the 18
  possibilities). The total area of the distributions have been
  rescaled to make them
  visible.]{\label{fig:corr}\includegraphics[width=7.3cm]{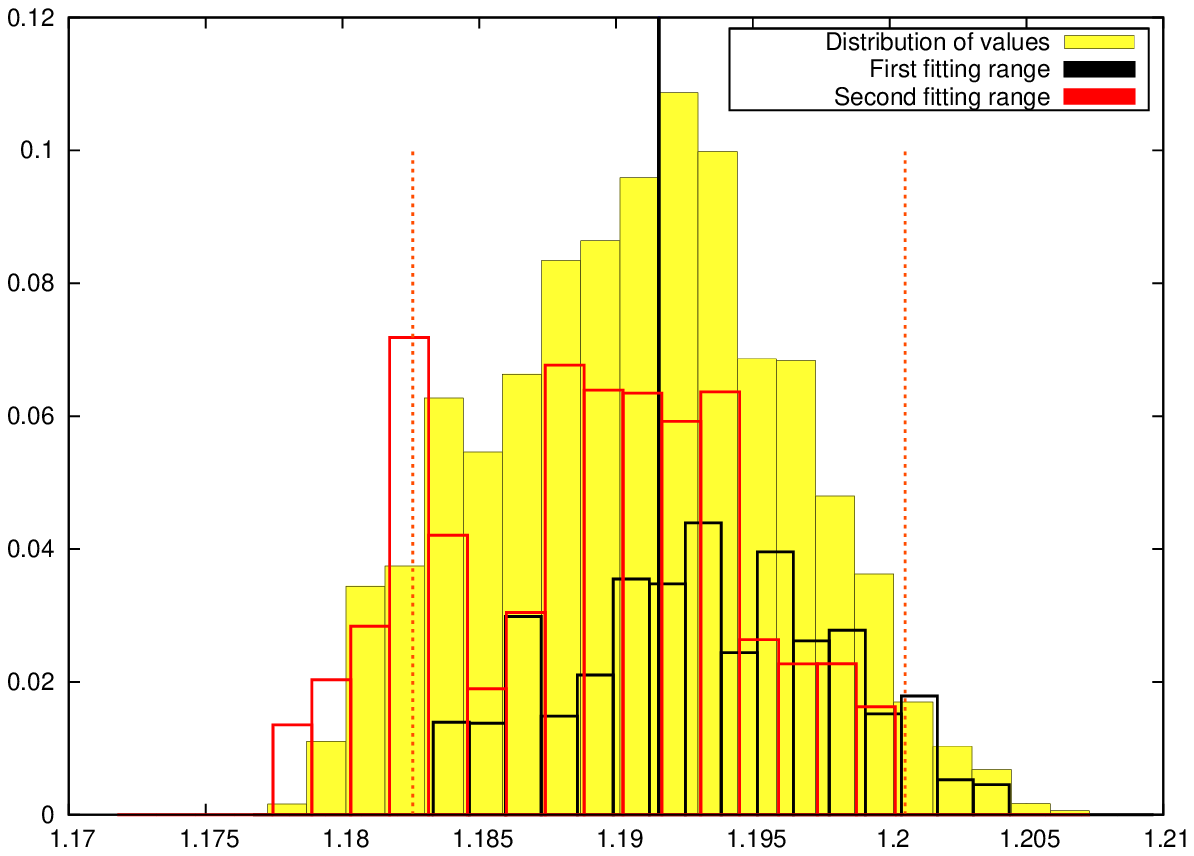}}\\ 
  \caption{Distributions with its contributions coming from different
    sources of theoretical error.}
\end{figure}

\begin{figure}[t,b]
  \centering
  \subfloat[Final distribution of values and the contribution to this
  corresponding to different scale settings.]{\label{fig:scale}\includegraphics[width=7.3cm]{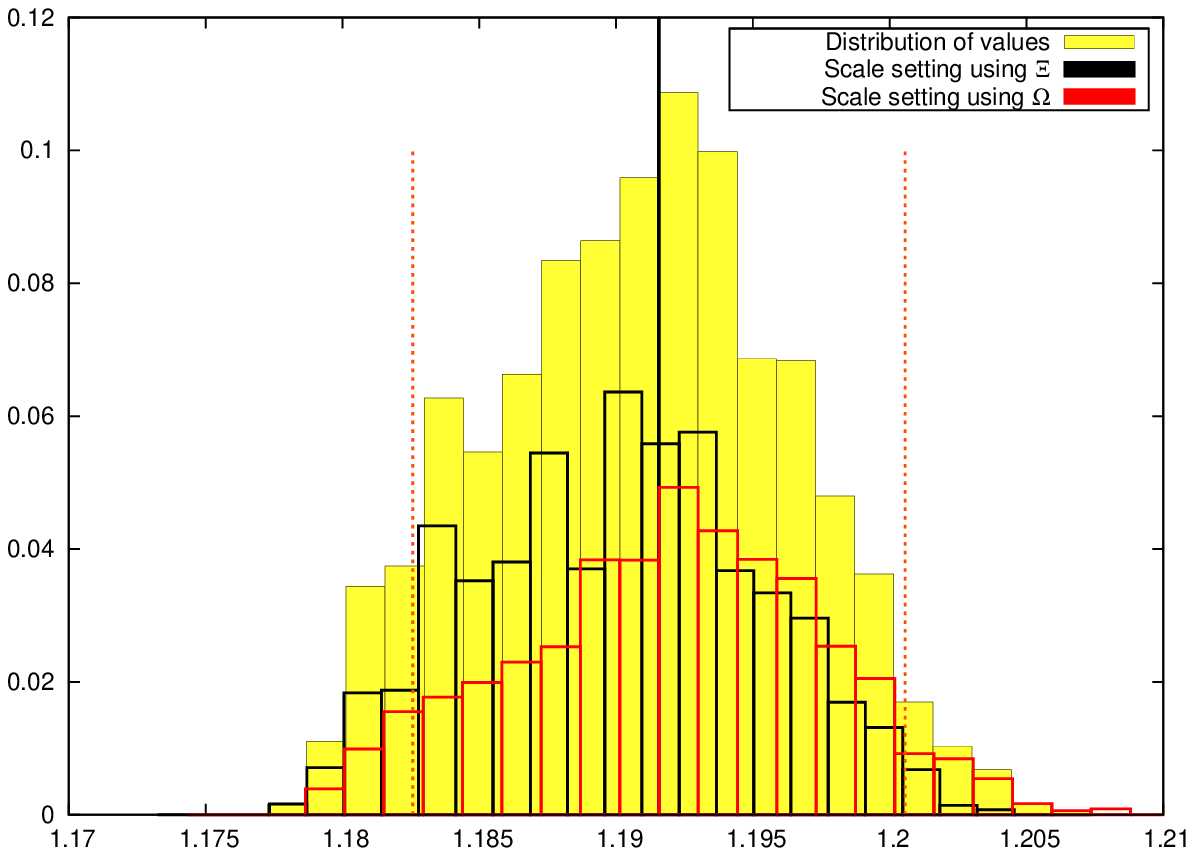}}\quad
  \subfloat[Final distribution of values, compared with the
  distributions corresponding to a different (1-loop and upper bound)
  finite volume correction.]{\label{fig:fv}\includegraphics[width=7.3cm]{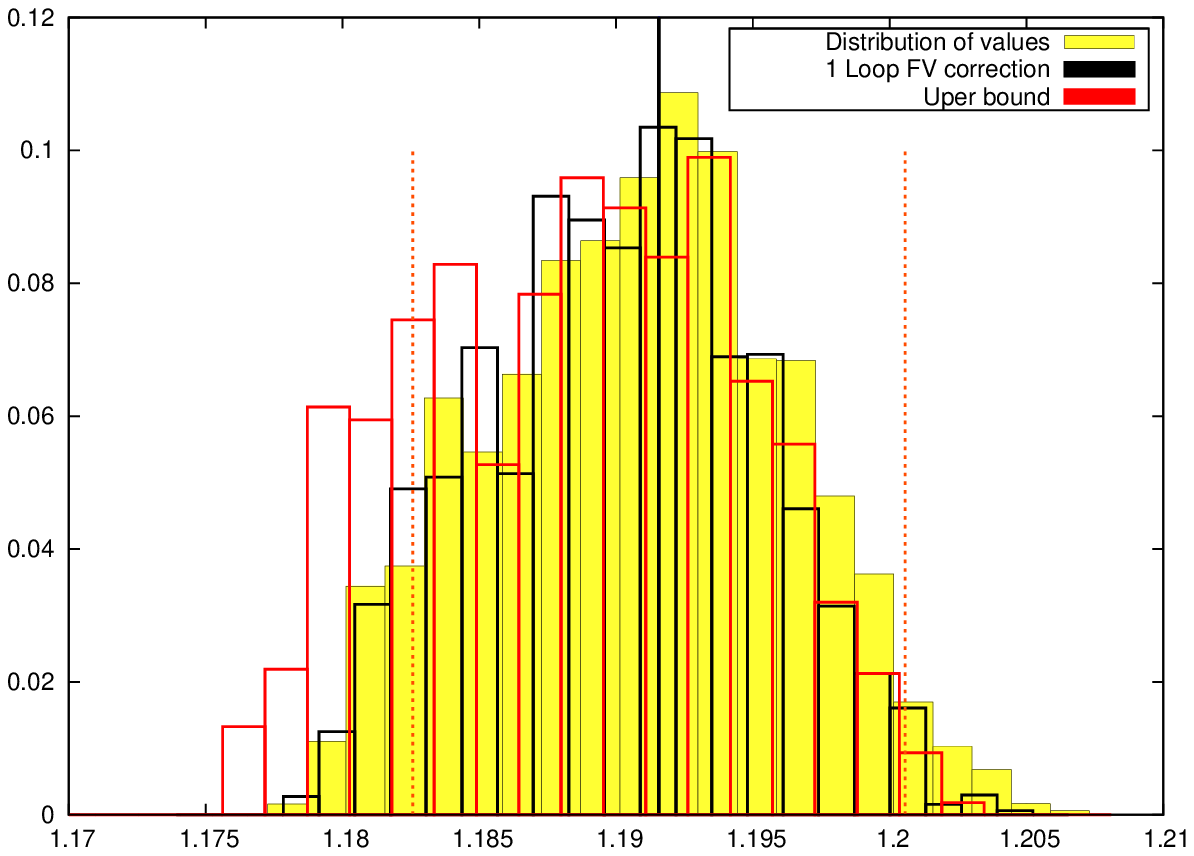}}\\
  \caption{Distributions with its contributions coming from different
    sources of theoretical error.}
\end{figure}

The next source of systematic error in importance are cutoff
effects. As was commented earlier, we have three possible
parametrizations for the cutoff effects: no cutoff effects at all, a
flavor breaking term proportional to $a$, and a flavor breaking term
proportional to $a^2$. In \fig{fig:cont} the
corresponding distributions are shown. In the same way \fig{fig:scale}
shows the contributions coming from the two possibilities for setting
the scale.

The two remaining sources of theoretical error deserve a separate
comment. First, the contamination with excited states is studied by
using a total of 18 different fitting ranges for the correlators,
corresponding to $t_\mr{min}/a=5$ or 6, for $\beta=3.3$; 7, 8 
or 9, for $\beta=3.57$; 10, 11 or 12 for $\beta=3.7$. In 
\fig{fig:corr} we can see the final distribution compared
with the distributions corresponding to $t_\mr{min}/a=5,7,10$ and
$t_\mr{min}/a=6,9,12$. These distributions have been rescaled (so
that they add to the total area of our final distribution), to make
the small distributions more visible.

Second, only the 2-loop finite volume corrections are included in the
1512 fits used to determine the final central value, because we do not
want to bias this value with the 1-loop contributions, that are, a
priori, less accurate than the 2-loop corrections. To estimate the
uncertainty associated with the subtraction of finite volume effects,
we repeated the full analysis with 1-loop finite volume corrections
and with an upper bound to the correction, computed by including only
the 2-loop correction to $F_\pi$. Our final distribution of values
with these two alternatives are plotted in \fig{fig:fv}. The error
associated with the finite volume effects is computed as the weighted
(by fit quality) standard deviation of the medians of these three
distributions, and 
added by quadratures to the 68\% confidence interval of our final
distribution to produce the final systematic error.

One final comment about the procedure to obtain the final systematic
error. The addition of different procedures to construct the final
distribution only can increase the systematic error. For example in
figure~(\ref{fig:mth}) we can clearly see that dropping \emph{any} of
the procedures to extrapolate to the physical point (for example not
using $SU(3)$ fits), gives a final value well in our final error band,
but a \emph{smaller} systematic error. This  general statement remains
true for the other sources of systematic error.

\acknowledgments

The author wants to thanks all the members of the
Budapest-Marseille-Wuppertal collaboration for their contributions to
the results presented here and S. D\"urr, C. Hoelbling and L. Lellouch
for their invaluable help writing this proceedings contribution.

\end{document}